\documentclass[aps,prl,twocolumn,showpacs,superscriptaddress,floatfix]{revtex4}
\usepackage{graphicx}
\usepackage{amssymb}

\begin{document}

\title{Superconductivity in the doped topological insulator Cu$_x$Bi$_2$Se$_3$ under high pressure}

\author{T. V. Bay} \affiliation{Van der Waals - Zeeman Institute, University of Amsterdam, Science Park 904, 1098 XH Amsterdam, The Netherlands}
\author{T. Naka} \affiliation{National Institute for Materials Science, Sengen 1-2-1, Tsukuba, Ibaraki 305-0047, Japan}
\author{Y. K. Huang} \affiliation{Van der Waals - Zeeman Institute, University of Amsterdam, Science Park 904, 1098 XH Amsterdam, The Netherlands}
\author{H. Luigjes} \affiliation{Van der Waals - Zeeman Institute, University of Amsterdam, Science Park 904, 1098 XH Amsterdam, The Netherlands}
\author{M. S. Golden} \affiliation{Van der Waals - Zeeman Institute, University of Amsterdam, Science Park 904, 1098 XH Amsterdam, The Netherlands}
\author{A. de Visser} \email{a.devisser@uva.nl} \affiliation{Van der Waals - Zeeman Institute, University of Amsterdam, Science Park 904, 1098 XH Amsterdam, The Netherlands}

\date{\today}

\begin{abstract} We report a high-pressure single crystal study of the topological superconductor Cu$_x$Bi$_2$Se$_3$. Resistivity measurements under pressure show superconductivity is depressed smoothly. At the same time the metallic behavior is gradually lost. The upper critical field data $B_{c2}(T)$ under pressure collapse onto a universal curve. The absence of Pauli limiting and the comparison of $B_{c2}(T)$ to a polar state function point to spin-triplet superconductivity, but an anisotropic spin-singlet state cannot be discarded completely.
\end{abstract}


\maketitle

Recently, topological insulators have sparked a wide research interest, because they offer a new playground for the realization of novel states of quantum matter~\cite{Hasan&Kane2010,Qi&Zhang2010}. In 3D topological insulators (TI) the bulk is insulating, but the 2D surface states $-$ protected by a non-trivial $Z_2$ topology $-$ are conducting. Exemplary TIs are the well-known thermoelectric effect materials Bi$_2$Te$_3$ and Bi$_2$Se$_3$, where photoemission ~\cite{Chen2009,Xia2009} and magnetotransport~\cite{Qu2010,Analytis2010} experiments have demonstrated the presence of a Dirac cone in the energy dispersion of the surface states, as in graphene, and 2D signatures like the quantum Hall effect.

The concept of topological insulators can also be applied to superconductors (SCs), due to the direct analogy between topological band theory and superconductivity: the Bogoliubov - de Gennes Hamiltonian for the quasiparticles of a SC has a close similarity to the Hamiltonian of a band insulator, where the SC gap corresponds to the gap of the band insulator ~\cite{Kitaev2009,Schnyder2009}. All topological SCs in 1, 2 and 3D are predicted to be non-trivial SCs with odd-parity Cooper pair states ~\cite{Sato2009,Sato2010}. Of major interest in the field of topological SCs is the realization of Majorana fermions~\cite{Hasan&Kane2010, Qi&Zhang2010}, that are predicted to exist as protected bound states on the edge of the 1, 2 or 3D superconductor. Majorana fermions are of great potential interest for topological quantum computation~\cite{Hasan&Kane2010, Qi&Zhang2010}. However, topological SCs are scarce. The B phase of $^3 $He has recently been identified as an odd-parity time-reversal invariant topological \textit{superfluid} (class 3D $Z$)~\cite{Qi2009}, whereas the correlated metal Sr$_2$RuO$_4$ is a time-reversal symmetry breaking chiral 2D $p$-wave SC (class D $Z$)~\cite{Kitaev2009}. Here we focus on a new potential candidate, namely the doped topological insulator Cu$_x$Bi$_2$Se$_3$~\cite{Hor2010}.

Recently, Hor and co-workers initiated a new route to fabricate topological superconductors, namely by reacting the TIs Bi$_2$Se$_3$ and Bi$_2$Te$_3$ with Cu or Pd ~\cite{Hor2010,Hor2011}. By intercalating Cu$^{1+}$ into the Van der Waals gaps between the Bi$_2$Se$_3$ layers, SC occurs with a transition temperature $T_c = 3.8$~K in Cu$_x$Bi$_2$Se$_3$ with $0.12 \leq  x \leq  0.15$. 
However, the reported SC shielding fractions were rather small and the resistance never attained a zero value below $T_c$, which cast some doubt on the bulk nature of SC. As regards Cu$_x$Bi$_2$Se$_3$, this concern was taken away by Kriener \textit{et al}.~\cite{Kriener2011a,Kriener2011b}. Bi$_2$Se$_3$ single crystals electrochemically intercalated by Cu$^{1+}$ have a SC volume fraction of about 60 \% (for $x=0.29$) as evidenced by the large jump in the electronic specific heat at $T_c$. SC was found to be robust and present for $0.1 \leq  x \leq  0.6$.

Photoemission experiments conducted to study the bulk and surface electron dynamics reveal that the topological character is preserved in Cu$_x$Bi$_2$Se$_3$~\cite{Wray2010}. Based on the topological invariants of the Fermi surface, Cu$_x$Bi$_2$Se$_3$ is expected to be a time-reversal invariant fully-gapped odd-parity topological SC~\cite{Sato2009,Sato2010}. A recent study of the Cooper pairing symmetry within a two-orbital model led to the proposal that such a state can be favored by strong-spin orbit coupling~\cite{Fu&Berg2010}. Indeed several experiments have revealed properties in line with topological SC. The magnetization in the SC state has an unusual field variation and shows curious relaxation phenomena, pointing to a spin-triplet vortex phase~\cite{Das2011}. Point contact measurements reveal a zero-bias conductance peak in the spectra, which possibly  provides a first evidence of Majorana fermions~\cite{Sasaki2011}. These signatures of topological SC make the experimental determination of the basic SC behavior of Cu$_x$Bi$_2$Se$_3$ highly relevant.

In this Letter we report the response of the superconducting phase of Cu$_{0.3}$Bi$_{2.1}$Se$_3$ to high pressure. SC is depressed smoothly. The $T$ variation of the upper critical field $B_{c2}(T)$ shows a universal behavior as a function of pressure, which lies above the standard variation for an $s$-wave SC. In addition, the absence of Pauli limiting and the $B_{c2}(T)$ variation in agreement with the polar state model both point to spin-triplet SC.

Single crystals of Cu$_x$Bi$_2$Se$_3$ were prepared by melting high purity elements at 850~$^\circ$C in sealed evacuated quartz tubes, followed by slowly cooling till 500-600~$^\circ$C. After growth the samples were annealed for 60-100 hours. The best samples $-$ with a zero resistance state below $T_c$ $-$ were obtained for a nominal Cu content $x \simeq 0.3$, a nominal Bi content $2.1$, and rapidly quenching after annealing. A typical resistivity trace for a current in the $ab$-plane of the rhombohedral crystal structure ($R \overline{3}
m$ spacegroup) is presented in the upper inset of Fig.~1. The resistivity shows metallic behavior, the residual resistance $\rho _0 = 0.15 ~$m$\Omega$cm is relatively low, and the carrier concentration $n \simeq 1.2 \times 10^{20}$~cm$^{-3}$. These transport parameters are similar to those reported in Refs.~\cite{Hor2010,Kriener2011a}.

\begin{figure}
\includegraphics[width=8cm]{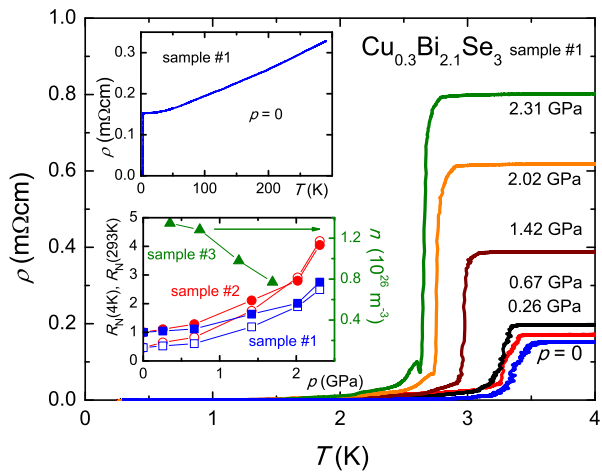}
\caption{(color online) Resistivity of Cu$_{0.3}$Bi$_{2.1}$Se$_3$ as a function of temperature around $T_c$ at pressures up to 2.31 GPa as indicated. Upper inset: $\rho (T)$ at $p=0$ for sample \#1. Lower
inset, left axis: $R_N$(293~K) (closed symbols) and $R_N$(4~K) (open symbols) as a function of pressure for samples \#1 and \#2. The subscript $N$ means the data are normalized at $R$(293~K,~$p=0$). Lower inset, right axis: (triangles) Carrier concentration $n(p)$ at $T = 4$~K for sample \#3. }
\end{figure}

The high-pressure measurements were carried out using a hybrid clamp cell made of NiCrAl and CuBe alloys. Samples were mounted on a plug which was placed in a Teflon cylinder with Daphne oil 7373 as hydrostatic pressure transmitting medium. The pressure cell was attached to the cold plate of a $^3$He refrigerator ($T_{base} = 0.24$~K). The ac-resistivity ($f = 13$~Hz) was measured, using the lock-in technique with a low excitation current ($I = 100 ~\mu$A), on two plate-like single-crystals with the current in the $ab$ plane. The suppression of SC in a magnetic field was investigated by resistance measurements in fixed fields $B \parallel ab$ (sample \#1) and $B \parallel c$ (sample \#2). The pressure was determined by measuring the SC transition temperature $T_c (p)$ of Pb in a separate experiment~\cite{Slooten2009}.

\begin{figure}
\includegraphics[width=6cm]{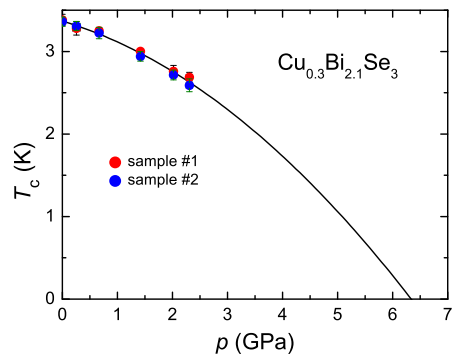}
\caption{(color online) Superconducting transition temperature as a function of pressure for samples \#1 and \#2 as indicated. The solid line represents a polynomial fit with linear and quadratic terms which serves to extrapolate the data.}
\end{figure}

In Fig.~1 we show $\rho (T)$ around $T_c$ for sample \#1 at pressures up to $2.31$~GPa. The normal state resistance $R$(4~K) shows a pronounced increase as a function of pressure, while $T_c$ steadily decreases. The overall good-quality of sample \#1 is attested by the relatively small width $\Delta T_c$ (as measured between 10~\% and 90~\% of $R$(4~K)), which ranges from $0.25$~K at $p=0$ to $0.06$~K at $p=2.31$~GPa. Nevertheless, a tail towards low temperatures associated with $\sim 10$~\% of the resistance path is present. For sample \#2 the resistance does not reach $R=0$, but remains finite at the level of 10 \% of $R$(4~K). In Fig.~2 we report the variation $T_c (p)$, determined by the midpoints of the resistive transitions, which almost coincide for both samples. The solid line in Fig.~2 (see caption) suggests $T_c$ might be suppressed at a critical pressure $p_c$ as high as $\sim 6.3$~GPa.

The depression of $T_c$ can be understood qualitatively in a simple model for a low carrier density SC where $T_c \sim \Theta_D \exp [-1/N(0)V_0]$, with $\Theta_D$ the Debye temperature, $N(0) \sim m^* n^{1/3}$ the density of states (with $m^*$ the effective mass) and $V_0$ the effective interaction parameter~\cite{Cohen1969}. The increase of $R$(4~K) under pressure by a factor $>$~5 indicates a decrease of the carrier concentration $n$, which in turn leads to a reduction of $N(0)$ and $T_c$. The reduction of $n$ is also apparent in $R (T)$ which gradually loses its metallic character (see lower inset in Fig~1): for $p \geq 2.02$~GPa $R$(4~K) exceeds $R$(293~K) for sample \#2. Hall effect data taken at $T=4~K$ on a third sample  (\#3) for $B~||~c$ confirm $n$ decreases under pressure (see lower inset in Fig.~1).

\begin{figure}
\includegraphics[width=7cm]{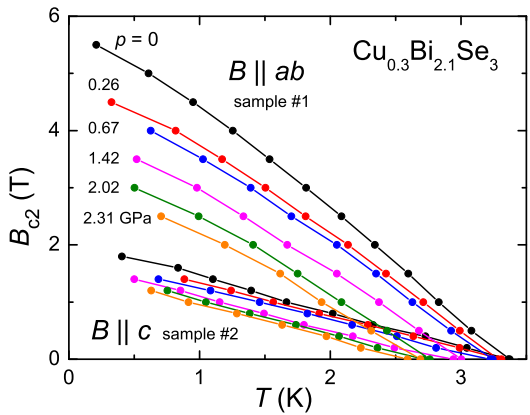}
\caption{(color online) Temperature variation of the upper critical field $B_{c2}(T)$ for $B \parallel ab$ and $B \parallel c$ at pressures of 0, 0.26, 0.67, 1.42, 2.02 and 2.31 GPa (from top to bottom).}
\end{figure}

\begin{figure}
\includegraphics[width=7cm]{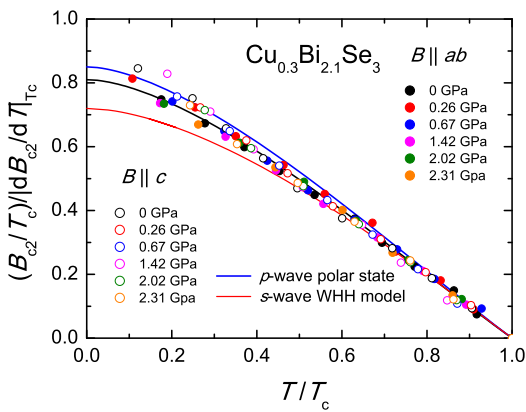}
\caption{(color online) Upper critical field $B_{c2}(T)$ divided by $T_c$ and normalized by the initial slope $|dB_{c2}/dT|_{T_c}$ as a function of the reduced temperature $T/T_c$ at pressures of 0, 0.26, 0.67, 1.42, 2.02 and 2.31 GPa for $B \parallel ab$ (closed symbols) and $B \parallel c$ (open symbols). The red and blue full lines represent model calculations for an $s$- and $p$-wave superconductor. The black curve matches the data closely and represents the polar state model function scaled by a factor 0.95, which would result from a 5 \% larger initial slope in the model.}
\end{figure}

Next we turn to the pressure variation of the upper critical field $B_{c2}(T)$. The depression of SC by a magnetic field was measured for $B \parallel ab$ ($B_{c2}^{ab}$~$-$~sample \#1) and $B \parallel c$ ($B_{c2}^{c}$~$-$~sample \#2). $T_c (B)$ extracted as the midpoints of the transitions is reported for each pressure in Fig.~3. We first discuss the results at $p=0$. The upper critical field shows a moderate anisotropy for a layered compound with $B_{c2}^{ab} (T\rightarrow 0) = 5.6$~T and $B_{c2}^{c} (T \rightarrow 0)= 1.9$ T. The anisotropy parameter $\gamma ^{an} = B_{c2}^{ab}/B_{c2}^{c} = 2.9$. Using the relations $B_{c2}^c = \Phi _0 / 2 \pi \xi _{ab} ^2$ and $B_{c2}^{ab} = \Phi _0 / 2 \pi \xi _{ab} \xi _{c}$, where $\Phi _0$ is the flux quantum, we derive SC coherence lengths $\xi _{ab} = 13$~nm and $\xi _{c}=4$~nm. These values are in good agreement with those reported recently~\cite{Hor2010,Kriener2011a}. It is important to distinguish whether our samples are in the clean or dirty limit, or in between. Notably, because a sufficiently clean sample with an electron mean free path $\ell$ larger than $\xi$ is a prerequisite for triplet Cooper pairing~\cite{Balian&Werthamer1963}. An estimate for $\ell$ can be obtained from the relation $\ell = \hbar k_F /\rho _0 n e^2$, assuming a spherical Fermi surface $S_F = 4 \pi k_F ^2$ with wave number $k_F =(3 \pi ^2 n)^{1/3}$. With $n = 1.2 \times 10^{26}$~m$^{-3}$ and $\rho _0 = 1.5 \times 10^{-6}~ \Omega$m (see Fig.~1) we calculate $k_F = 1.5 \times 10^9$~m$^{-1}$ and $\ell = 34$~nm, which ensures $\ell > \xi$.

A more detailed analysis can be made employing the slope of the upper critical field $dB_{c2}/dT$ at $T_c$~\cite{Orlando1979}. In the dirty limit the initial slope is given by $|dB_{c2}/dT|_{T_c} =~4480 \cdot \gamma \rho_0 $, where $\gamma$ is the Sommerfeld coefficient of the specific heat. With $\gamma =$  22.9~J/m$^3$K$^2$~\cite{Kriener2011a} we calculate $|dB_{c2}/dT|_{T_c} = 0.15$~T/K. This value is much lower than the measured values 2.0 T/K ($B \parallel ab$) and 0.6 T/K ($B \parallel c$), which confirms our samples are not in the dirty limit. By adding the clean limit term $|dB_{c2}/dT|_{T_c} =~1.38 \times 10^{35} \cdot \gamma ^2 T_c / S_F ^2$ in the model~\cite{Orlando1979} estimates for $\ell$ and $\xi$ can be extracted from the experimental values of $|dB_{c2}/dT|_{T_c}$. For $B \parallel ab ~(c)$ we obtain $\ell \sim 90~ (45)$~ nm and $\xi \sim 9 ~(19)$~nm. Notice, in this analysis we used the normal-state $\gamma$-value as input parameter. If we consider that only part of the sample becomes superconducting, a reduced $\gamma _s$ value~\cite{Kriener2011a} should be used. This will affect the absolute values of the deduced parameters, but not our conclusion $\ell > \xi$. We conclude our samples are sufficiently pure to allow for odd-parity SC.

Under pressure $B_{c2}(T)$ gradually decreases and the anisotropy parameter reduces from $\gamma ^{an} = 2.9$ at $p=0$ to 2.1 at the highest pressure $p= 2.31$~GPa. The coherence lengths increase to $\xi _{ab} = 15$~nm and $\xi _{c}=7$~nm. As mentioned above the increase of $\rho _0 $ and the gradual loss of metallic behavior under pressure can for the major part be attributed to a corresponding decrease of $n$, which tells us the ratio $\ell / \xi > 1$ in the entire pressure range.

The functional behavior of $B_{c2}$ does not change with pressure, as is demonstrated in Fig.~4. All the $B_{c2}(T)$ curves collapse on one single universal function $b^*(t)$, with $b^* = (B_{c2}/T_{c})/|dB_{c2}/dT|_{T_{c}}$ and $t = T/T_c$ the reduced temperature. This holds for $B \parallel ab$, as well as for $B \parallel c$. In order to analyze the data further we have traced in Fig.~4 also the universal curve for a clean spin-singlet SC with orbital limited upper critical field $B_{c2}^{orb}(0)=0.72 \times T_c ~|dB_{c2}/dT|_{T_{c}}$ (WHH model~\cite{Werthamer1966}), noting that for dirty limit system the prefactor would reduce to $0.69$. Clearly, the data deviate from the standard spin-singlet behavior. Next, we consider the suppression of the spin-singlet state by  paramagnetic limiting~\cite{Clogston1962,Chandrasekhar1962}. The Pauli limiting field in the case of weak coupling is given by $B^P (0) = 1.86 \times T_c$. For Cu$_{0.3}$Bi$_{2.1}$Se$_3$ $B^P (0) = 6.2~T$ at ambient pressure. When both orbital and spin limiting fields are present, the resulting critical field is $B_{c2}(0)=B_{c2}^{orb}(0)/\sqrt{1 + \alpha ^2}$, with the Maki parameter $\alpha = \sqrt {2} B_{c2}^{orb}(0)/ B^P (0)$~\cite{Werthamer1966,Maki1966}. For $B \parallel ab~(c)~$ we calculate $\alpha = 1.1~(0.34)$ and $B_{c2}(0) = 3.3~(1.4)$~T. These values of $B_{c2} (0)$ are much lower than the experimental values (see Fig.~3) and we conclude the effect of Pauli limiting is absent. In general, by including the effect of paramagnetic limiting
the overall critical field is reduced to below the universal spin-singlet values~\cite{Werthamer1966,Orlando1979}. Thus, the fact that our $B_{c2}$ data are well above even these universal values points to an absence of Pauli limiting, and is a strong argument in favor of spin-triplet SC. The Pauli paramagnetic effect suppresses spin-singlet Cooper pairing, as well as the $L_z = 0$ triplet component $(\mid \uparrow \downarrow \rangle + \mid \downarrow \uparrow \rangle )/ \sqrt 2$, while the equal-spin pairing (ESP) states $\mid \uparrow \uparrow \rangle$ and $\mid \downarrow \downarrow \rangle$ with $L_z =1$ and $L_z=-1$, respectively, are stabilized in a high magnetic field. Exemplary SCs where Pauli limiting is absent are the spin-triplet SC ferromagnets URhGe~\cite{Hardy&Huxley2005} and UCoGe~\cite{Huy2008}.

Next we consider the role of anisotropy of the crystal structure. Calculations show that for layered SCs, for $B$ parallel to the layers, $B_{c2}^{orb}$ is reduced and the critical field can exceed the values of the WHH model~\cite{Klemm1975}. Cu$_x$Bi$_2$Se$_3$ is a layered compound~\cite{Hor2010} with a moderate anisotropy $\gamma ^{an} = 2.9$. In this respect it is interesting to compare to other layered SCs, like alkali intercalates of the semiconductor MoS$_2$~\cite{Woollam&Somoano1976}, which have $\gamma ^{an}$-values in the range $3.2-6.7$. A striking experimental property of these layered SCs is a pronounced upward curvature of $B_{c2}(T)$ for $B$ parallel to the layers for $T < T_c$ due to dimensional cross-over, which is also a salient feature of model calculations~\cite{Klemm1975}. However, an upward curvature is not observed in Cu$_x$Bi$_2$Se$_3$. In more detailed theoretical work the anisotropy of both the Fermi surface and the SC pairing interaction has been incorporated~\cite{Youngner&Klemm1980}. Under certain conditions this can give rise to deviations above the WHH curve as seen in Fig.~4 above. The bulk conduction band Fermi surface of the parent material $n$-type Bi$_2$Se$_3$ is an ellipsoid of revolution along the $k_c$ axis with trigonal warping~\cite{Kohler1973}. Assuming Cu$_x$Bi$_2$Se$_3$ has a similarly shaped Fermi surface, the anisotropy would result in a different functional dependence of $B_{c2}(T)$ for $B \parallel ab$ and $B \parallel c$. On the qualitative level this is at variance with the universal $b^* (t)$ reported in Fig.~4.

Finally, we compare the $B_{c2}(T)$ data with upper critical field calculations for a $p$-wave SC~\cite{Scharnberg&Klemm1980}. For an isotropic $p$-wave interaction the polar state (which applies for a linear combination of both ESP components) has the highest critical field for all directions of the magnetic field. In Fig.~4 we compare the $B_{c2}(T)$ data with the polar state model function. This time the data lie below the model curve, but most importantly, the temperature variation itself is in agreement with the model, as illustrated by the solid black curve in Fig.~4. We have also considered a scaled WHH curve, but it fits the data much less well: increasing $b^* (0)$ by $e.g.$ $10~\%$ to match the experimental value, results in an overall curvature of $b^* (t)$ in disaccord with the data. In general, topological SC involves all three components of the triplet state with a full gap in zero field~\cite{Sato2009,Sato2010}. In the case of Cu$_x$Bi$_2$Se$_3$ model calculations~\cite{Fu&Berg2010} indicate triplet pairing is possibly restricted to the $L_z =0$ component. In an applied magnetic field we expect a phase transition or cross-over to a polar state to occur. Clearly, more theoretical work is needed on topological superconductors in a magnetic field to settle the issue of $B_{c2}$.

In summary, we have investigated the pressure variation of the superconducting phase induced by Cu intercalation of the topological insulator Bi$_{2}$Se$_3$. Superconductivity is robust and by extrapolating $T_c (p)$ appears to vanish at the high critical pressure of $p_c = 6.3$~GPa. The metallic behavior is gradually lost under pressure. The upper-critical field $B_{c2}$ data under pressure collapse onto a single universal curve, which differs from the standard curve of a weak-coupling, orbital-limited, spin-singlet superconductor. The absence of Pauli limiting, the sufficiently large mean free path, and the polar-state temperature variation of $B_{c2}$ data, point to Cu$_x$Bi$_{2.1}$Se$_3$ as a $p$-wave superconductor.

Acknowledgements $-$ T.V. Bay acknowledges support of the Vietnamese Ministry of Education and Training. The authors are grateful to S. Ramakrishnan, B.P. Joshi and R. A. Klemm for fruitful discussions.

\bibliography{RefsTI}

\end{document}